\documentclass[10pt,journal]{IEEEtran} 
\IEEEoverridecommandlockouts

\usepackage{amsfonts, amsmath,amssymb,amsthm}
\usepackage{algorithm, algorithmic}
\usepackage{array}
\usepackage{cases}
\usepackage{changepage}
\usepackage[compress]{cite}
\usepackage{epstopdf}
\usepackage[english]{babel}
\usepackage{etoolbox}
\usepackage{graphicx}
\usepackage{multirow}
\usepackage{physics} 	
\usepackage{subfigure}
\usepackage{setspace}
\usepackage{url}  
\usepackage[hyphenbreaks]{breakurl} 
\usepackage{lipsum}
\usepackage{nicefrac}	
\usepackage[table,xcdraw]{xcolor}	
\usepackage{textcomp}
\usepackage{makecell}
\usepackage[font=small]{caption} 
\usepackage{mathrsfs}
\usepackage{indentfirst} 
\usepackage{textcomp} 

\def\ps@headings{%
\def\@oddhead{\mbox{}\scriptsize\rightmark \hfil \thepage}%
\def\@evenhead{\scriptsize\thepage \hfil \leftmark\mbox{}}%
\def\@oddfoot{}%
\def\@evenfoot{}}
\makeatother
\makeatletter
\patchcmd{\@thm}
  {\trivlist}
  {\list{}{\leftmargin=\thm@leftmargin\rightmargin=\thm@rightmargin}}
  {}{}
\patchcmd{\@endtheorem}
  {\endtrivlist}
  {\endlist}
  {}{}
\newlength{\thm@leftmargin}
\newlength{\thm@rightmargin}

\newcommand{\xnewtheorem}[3]{%
  \newenvironment{#3}
    {\thm@leftmargin=#1\relax\thm@rightmargin=#2\relax\begin{#3INNER}}
    {\end{#3INNER}}%
  \newtheorem{#3INNER}%
}

\makeatother

\newtheoremstyle{indentedupright}{3pt}{3pt}{} {}{\bfseries}{.}{.5em}{} 
\newtheoremstyle{indenteditalic}{3pt}{3pt}{\itshape} {}{\bfseries}{.}{.5em}{} 

\theoremstyle{indenteditalic}
\xnewtheorem{0pt}{0pt}{oldpro}{Conventional Property}
\xnewtheorem{0pt}{0pt}{newpro}{New Property}
\xnewtheorem{0pt}{0pt}{obser}{Observation}
\xnewtheorem{0pt}{0pt}{pro}{Property}
\xnewtheorem{0pt}{0pt}{lem}{Lemma}
\xnewtheorem{0pt}{0pt}{corl}{Corollary}
\xnewtheorem{0pt}{0pt}{addcon}{Additional Constraint}

\newcommand{\romu}[1]{\uppercase\expandafter{\romannumeral #1\relax}} 
\newcommand{\roml}[1]{\lowercase\expandafter{\romannumeral #1\relax}}    

\renewcommand{\baselinestretch}{1.0}

\begin{document}
\title{\LARGE Underwater Acoustic Reconfigurable Intelligent Surfaces: from Principle to Practice}
\author{\IEEEauthorblockN{Yu Luo\IEEEauthorrefmark{1}, Lina Pu\IEEEauthorrefmark{2}, Junming Diao\IEEEauthorrefmark{1}}, Chun-Hung Liu\IEEEauthorrefmark{1}, Aijun Song\IEEEauthorrefmark{3}\\
\IEEEauthorblockA{\IEEEauthorrefmark{1}ECE Department, Mississippi State University, Mississippi State, MS, 39759, USA\\
\IEEEauthorrefmark{2}Department of Computer Science, University of Alabama, Tuscaloosa, AL 35487, USA\\
\IEEEauthorrefmark{3}Department of Electrical and Computer Engineering, University of Alabama, Tuscaloosa, AL 35487, USA\\
Email: \{yu.luo, jdiao, chliu\}@ece.msstate.edu, lina.pu@ua.edu, song@eng.ua.edu}
}

\maketitle

\begin{abstract}
\label{sec:Abs}
This article explores the potential of underwater acoustic reconfigurable intelligent surfaces (UA-RIS) for facilitating long-range and eco-friendly communication in marine environments. Unlike radio frequency-based RIS (RF-RIS), which have been extensively investigated in terrestrial contexts, UA-RIS is an emerging field of study. The distinct characteristics of acoustic waves, including their slow propagation speed and potential for noise pollution affecting marine life, necessitate a fundamentally different approach to the architecture and design principles of UA-RIS compared to RF-RIS. Currently, there is a scarcity of real systems and experimental data to validate the feasibility of UA-RIS in practical applications. To fill this gap, this article presents field tests conducted with a prototype UA-RIS consisting of 24 acoustic elements. The results demonstrate that the developed prototype can effectively reflect acoustic waves to any specified directions through passive beamforming, thereby substantially extending the range and data rate of underwater communication systems.
\end{abstract}

\section{Introduction}
\label{sec:Int}

In underwater environments, the propagation of RF and optical signals is severely restricted due to significant absorption attenuation and scattering. As a result, acoustic communication has become the principal method for long-range communication in marine settings.

Despite considerable research over the past decades, extending the communication range for high data rate acoustic communication in underwater settings has proven challenging~\cite{zhu2023internet}. While lower frequency acoustic signals can achieve extremely long ranges due to their lower propagation losses, high data rate transmissions require higher frequencies, which are subject to significant propagation losses. Additionally, the acoustic communication is constrained by limited battery capacity and the need to moderate transmission power to mitigate the impact on marine wildlife from anthropogenic acoustic noise~\cite{duarte2021soundscape}. Consequently, developing high-rate, long-range, and environmentally sustainable underwater acoustic communication systems remains a challenging task.

The success of radio frequency reconfigurable intelligent surfaces \textcolor{black}{(RF-RIS)} in terrestrial communications has sparked interest in their potential to enhance underwater acoustic\,(UA) systems as well~\cite{hassouna2023survey, liu2021reconfigurable}. By strategically manipulating the amplitude and phase of reflected waves, UA-RIS technology can align the direct and reflected waves to superimpose at the receiver. This approach effectively mitigates propagation loss and significantly \textcolor{black}{boosts} signal strength, even at high frequencies Additionally, with the assistance of \textcolor{black}{UA-RIS, senders} can transmit data at lower sound levels, reducing their environmental footprint while extending battery life.

Although there are many attractive features, current research on RIS primarily focuses on terrestrial scenarios using radios as signal carriers~\cite{ding2022state}. Unfortunately, the physical properties of acoustic waves are completely different from those of RF signals, rendering existing RF-RIS designs unsuitable for direct application in underwater environments. At present, only a handful of studies explore the design of UA-RIS~\cite{sun2022high, wang2023designing}. However, these works evaluate the performance of the proposed system through simulations and theoretical analysis, lacking experimental validation to confirm the efficiency of the proposed UA-RIS architecture.
To bridge the gap between theoretical frameworks and practical applications, \textcolor{black}{this work introduces several key advancements in UA-RIS research.}

\textcolor{black}{While RF-RIS typically employs varactor diodes and tunable inductors for precise phase manipulation, which have been adapted in existing conceptual designs for UA-RIS~\cite{sun2022high, wang2023designing}, these varactor-based systems are inadequate for underwater environments. The tuning range of varactors (nH for inductors and pF for capacitors) is significantly lower than what is required for UA-RIS (mH for inductances and nF for capacitances). To address this, we adopt a fundamentally different UA-RIS architecture featuring a load network.}


We developed a 1-bit phase coding based UA-RIS platform consisting of 24 (6 $\!\!\times\!\!$ 4) reflection units. Each unit is capable of independently switching the phase of reflected acoustic waves between 0$^{\circ}$ (in-phase) and 180$^{\circ}$ (antiphase) through coding the load impedance of the reflection units using a microcontroller (MCU). By configuring the coding schemes of the unit reflectors, the reflected waves will superpose in the specified direction, thereby enhancing the signal strength at the receiver side.

To empirically validate the effectiveness of our UA-RIS design, we conducted both tank and lake tests. These experiments demonstrated that our UA-RIS can flexibly enhance or attenuate the strength of acoustic waves in the desired direction, confirming the potential of UA-RIS to enable high-rate, long-range and environmentally friendly underwater communications.



\section{Potential Applications of UA-RIS}
\label{sec:App}
\textcolor{black}{ UA-RIS is emerging as a revolutionary technology enabling high-rate, long-range communications for autonomous underwater vehicles (AUVs), empowering environment-friendly channel sharing, and creating an alternative solution for secure acoustic communication.}

\subsection {\textcolor{black}{High-Rate, Long-Range Communications for AUV Network}}
\label{sec:frx }

\textcolor{black}{AUV networks for applications such as high-resolution image and video transfer in underwater inspection, cooperative exploration in AUV swarms over a large area, and extensive data exchange in future AI applications  require communication technologies that deliver both high rate and long-range capabilities~\cite{okereke2023overview}. However, achieving both attributes simultaneously is challenging due to the inherent size and weight constraints of AUVs. While Multiple-Input and Multiple-Output (MIMO) with compact transducers can facilitate high data rates~\cite{tanveer2023cross}, they suffer from limited range due to severe attenuation of high-frequency acoustic signals in ocean and the limited transmission power constrained by AUV battery capacity. Alternatively, long-range communication with lower central frequency typically requires large transducer arrays, infeasible for AUV integration due to their substantial size and weight.}

\begin{figure}[htb]
	\centerline{\includegraphics[width=8.5cm]{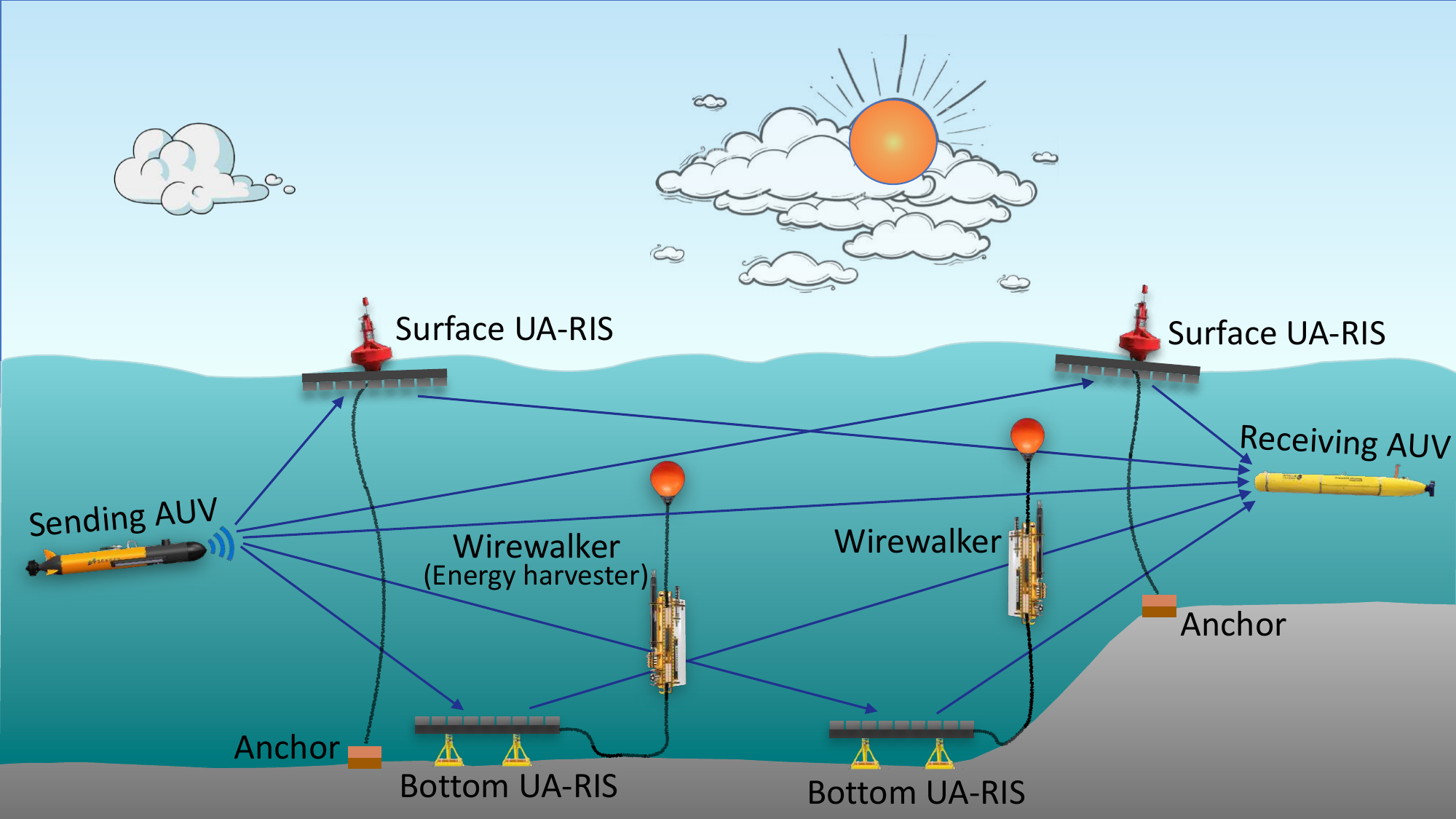}}
  	\caption[Caption for LOF]{Long-term vision of high-rate, long-range communication with UA-RIS, where surface UA-RIS  can be solar powered and seabed-mounted systems leverage energy harvesters like Wirewalker\footnotemark.}
  	\label{fig:Scenario}
\end{figure}
\footnotetext{Surface UA-RISs encounter significant mobility challenges arising from ocean waves. Therefore, we focus on seabed mounted UA-RISs in this work. Addressing the mobility issues associated with surface UA-RISs will be investigated in our future research. }

\textcolor{black}{To address these challenges, UA-RIS emerges as an enabling technology to overcome these limitations, enhancing AUV communication capabilities without requiring modifications to the AUVs themselves by leveraging UA-RIS as infrastructure.} As illustrated in Fig.~\!\ref{fig:Scenario}, by intelligently modulating the phase and amplitude of these scattered waves, the UA-RISs can effectively steer all reflections toward the target destination. This process allows the receiver to markedly improve the signal-to-noise ratio (SNR) by combining multiple copies of the data signal bounced back by the UA-RISs, thereby significantly extending the communication range, even at high data rates.

\subsection {Environmental-Friendly Channel Sharing}
\label{sec:EnvFri}

Marine wildlife, which heavily relies on acoustic signals for communication, hunting, and navigation, \textcolor{black}{often operates within the same frequency ranges used by artificial acoustic systems}~\cite{luo2014challenges}. For instance, toothed whales typically communicate around 10 kHz, while bottlenose dolphins use frequencies ranging from 200\,Hz to 24\,kHz for whistle signals and 200\,Hz to 150\,kHz for echolocation clicks. Additionally, killer whales primarily employ echolocation signals within the 12\,kHz to 25\,kHz frequency band~\cite{richardson2013marine}.

\begin{figure}[htb]
\centerline{\includegraphics[width=8.7cm]{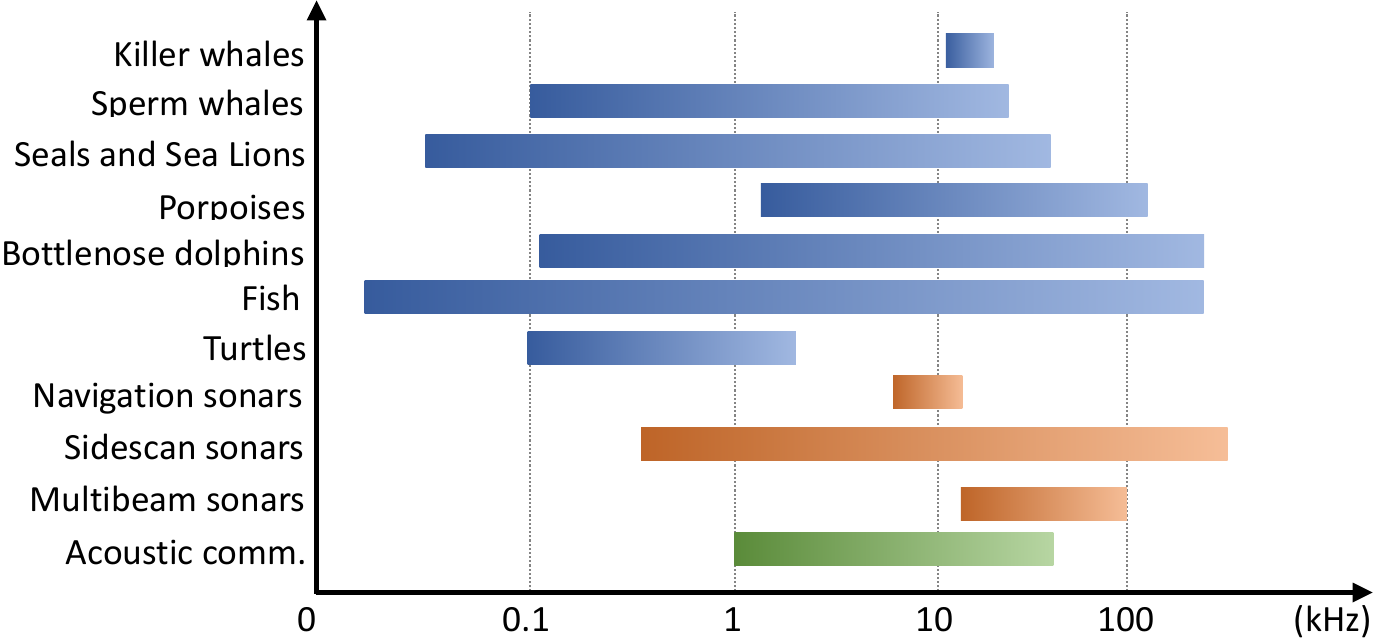}}
  \caption{Typical bandwidth usage for each animal group and artificial acoustic systems.}\label{fig:SpecShar}  
\end{figure}

Fig.~\!\ref{fig:SpecShar} illustrates the bandwidths of \textcolor{black}{different marine species, indicating significant overlaps with frequencies used by artificial systems}. This overlap raises concerns about possible disruptions or damage to the auditory systems of marine life, especially given the high sound levels typically involved in data transmissions. \textcolor{black}{UA-RIS addresses these challenges by} enabling data transmission at lower sound levels, which can mitigate the impact on the marine soundscape and enhance the protection of marine life. Additionally, UA-RIS systems direct reflected signals specifically, rather than dispersing them broadly, further reducing interference and minimizing the affected area. \textcolor{black}{This targeted approach is crucial for mitigating the anthropocene ocean soundscape and protecting marine ecosystems.}

\subsection {Enhanced Security for Acoustic Communication}
\label{sec:EnSec}
\textcolor{black}{In wireless communication systems, securing data transmissions generally involves the communicating parties} sharing a cryptographic key, which is used to encrypt and decrypt messages. However, underwater environments pose unique challenges for secure communications due to significant propagation delays and low data rates associated with acoustic signals. These factors make the process of establishing a secure communication channel particularly time-consuming, affecting the overall efficiency of secure data transmissions in underwater networks.

\textcolor{black}{UA-RIS creates new opportunities} to enhance information security at the physical layer.  By leveraging the reflective array, the transmitter is able to send data using lower power levels, preventing distant eavesdroppers from intercepting the signal with sufficient SNR.
UA-RIS can further bolster acoustic communication security by incorporating sophisticated array processing technologies. When the position of a potential eavesdropper is known, the minimum variance distortionless response (MVDR) algorithm \textcolor{black}{can be used to direct the reflected signal towards} the intended receiver, while simultaneously reducing its intensity towards the eavesdropper's location. This targeted suppression minimizes the risk of information leakage, thereby significantly enhancing the security of the acoustic communication channel.

\section{Challenging Issues of UA-RIS Design}
\label{sec:Challenge}
The exploration of RIS in underwater environments represent an entirely new field of study. The findings from research on RF-RIS cannot be directly applied to UA-RIS due to the distinct characteristics of acoustic waves and acoustic reflectors. In this section, we will delve into various challenges that arise in the development of UA-RIS.

\textbf{1) Different physics of acoustic wave:} The reflective units in UA-RIS are required to interact with acoustic signals, which are essential mechanical waves that exhibit physics entirely different from RF signals. Consequently, existing metamaterials and metasurfaces, originally devised for radio applications, are not working in aquatic environments. Addressing this issue necessitates the development of new types of reflecting units and associated architectures specifically tailored for UA-RIS applications.

\textbf{2) Low frequencies of acoustic signal:} RF-RIS systems are typically designed to reflect radio communication signals ranging from hundreds of megahertz to tens of gigahertz. In such systems, the phase of the reflected waves can be controlled by altering the length of transmission line or by adjusting the load impedance of the reflection unit using varactor diodes~\cite{boccia2002application}. In contrast, the operating frequency of UA-RIS is significantly lower, generally below 200\,kHz due to the substantial frequency-dependent attenuation of acoustic waves in water. Consequently, the wavelength of the received signal is much longer than the length of transmission line, and the load impedance provided by the varactor is insufficient to change the phase of reflected waves. \textcolor{black}{These factors necessitate new solutions for precise  phase manipulation in UA-RIS.}

\textbf{3) Limited \& unreliable power supply:}
RF-RIS systems are generally designed for installation on building walls, enabling straightforward connections to the power grid for a reliable energy supply. Conversely, UA-RIS systems operate in maritime settings, where a stable power infrastructure is often absent. Consequently, UA-RIS units, especially those situated at the ocean's floor, require energy harvesting from sources such as ocean waves or microorganisms for self-sufficiency~\cite{kelly2023prototyping}. Nonetheless, the power density provided by these sustainable sources tends to be low and unstable, presenting substantial challenges for UA-RIS regarding power consumption constraint. Addressing these limitations necessitates meticulous planning in both system design and power management strategy.

\textbf{4) High dynamic range of signal intensity:} Due to the substantial loss caused by viscous absorption, the attenuation of acoustic waves at medium and high frequencies is significantly higher than that experienced by radio signals in air. To counteract the high attenuation over long distances, acoustic transmitters are engineered to have high source levels, capable of achieving 180\,dB re 1\,$\mu$Pa\,@\,1\,m, or even higher. Consequently, the intensity of acoustic waves that reach the UA-RIS varies considerably depending on the distance from the source. The high dynamic range of the incident waves presents a significant challenge for the design of the UA-RIS. The system must accommodate both weak and strong signals, which necessitate different load network structures to accurately reflect waves with the desired amplitude and phase.

\textbf{5) Wideband reflection:} RF signals are typically modulated using high-frequency carrier waves to allow for smaller antenna sizes. Therefore, radio signals are usually narrowband, which is contrasts with acoustic signals. Specifically, the bandwidth of acoustic signals, spanning several kilohertz, is on the same order of magnitude as their central frequency. Consequently, UA-RIS must accommodate the frequency-dependent aspects of signal propagation and array response. This necessitates the use of advanced wideband array processing algorithms, demanding each reflecting unit in the UA-RIS to manipulate not just the phase but also the amplitude of the reflected wave.

\textbf{6) Interference concern:} As outlined in Section~\!\ref{sec:EnvFri}, artificial acoustic systems and marine wildlife share a similar frequency band for communication. As a result, acoustic interference can exert a greater effect on marine ecosystems compared to the impact of RF signals in terrestrial environments. In order to lessen this impact, it is crucial for the UA-RIS to minimize side-lobes in its reflected beam while ensuring high directional gain. Achieving this balance calls for a precise and thoughtful consideration of both the hardware design and the array processing algorithm in the UA-RIS.

\textbf{7) Size and weight constraint:} The structure of a typical RF-RIS can be considered as two-dimensional, making it possible to build massive reflecting units on a lightweight printed circuit board (PCB). In contrast, UA-RIS are inherently three-dimensional, with the volume of each reflector inversely proportional to its resonant frequency. Due to the lower frequency of communication signals, acoustic reflectors tend to be considerably larger and heavier. This imposes a practical limit on the number of reflecting units that can be integrated into a UA-RIS. Consequently, many RF-RIS designs that depend on large arrays for controlling wave reflection patterns may not work effectively for UA-RIS applications. Therefore, UA-RIS calls for innovative approaches to enhance the direction gain with limited reflection units.

\textbf{8) Low propagation speed:} In RF-RIS systems, the swift speed of RF signals ensures that both reflected and direct waves reach the receiver almost simultaneously. However, this contrasts with acoustic wave in water, which propagate at a much slower speed of only 1480\,m/s --- five orders of magnitude lower than RF signals in air. As a result, in applications of UA-RIS, the direct and reflected waves are likely to arrive the destination at different times. In this situation, the receiver obtains multiple copies of the raw signal reflected by the RIS. How to effectively combine these copies to improve the receiving SNR is a challenging issue.

\textcolor{black}{In this work, we focus on tackling the first three challenges. Instead of using varactor-based phase manipulation -- which is ineffective at acoustic frequencies due to insufficient tuning ranges -- our proposed UA-RIS achieves flexible phase shifts as well as high energy efficiency through a load network. The remaining challenges require new hardware designs, architectures, and algorithms, which we will explore in future research.}

\section{Architecture of UA-RIS}
\label{sec:Hardware}
In this section, we begin by examining the differences between UA-RIS units and conventional acoustic transducers. Next, we introduce the architecture of our 1-bit phase coding UA-RIS. Subsequently, we analyze the beam patterns of reflected waves with various coding schemes.

\subsection {Acoustic Transducers versus UA-RIS Reflection Units}
\label{sec:Aco}
In underwater environments, the requirements for acoustic communication and intelligent reflecting surfaces are distinctly different. Consequently, an acoustic transducer tailored for underwater communication is not directly suitable for constructing an intelligent surface.

In the context of acoustic communication, high sensitivity is crucial for the transducer to effectively convert energy between electrical signals and acoustic waves. On the other hand, for a UA-RIS, the priority is on efficiently reflecting signals, \textcolor{black}{making transducer sensitivity less critical}.

Additionally, acoustic communication typically requires the ability to transmit and receive signals omnidirectionally, necessitating the use of  transducers designed in cylindrical-shaped piezoelectric (PZT) (as depicted in the left corner of Fig.~\!\ref{fig:Cir}). In contrast, \textcolor{black}{reflection units in UA-RIS are densely deployed} and focus primarily on reflecting incident waves from the front, making signals from the side and rear negligible. \textcolor{black}{Thus, flat PZT pads, which emphasize frontal reflection, are more appropriate for UA-RIS.}

Furthermore, the high sensitivity and precision of commercial acoustic transducers come with a high cost. This poses a challenge for their integration into a UA-RIS, which generally necessitates a substantial number of unit cells, ranging from dozens to hundreds, to attain adequate reflection gain for long-range communication. 

\textcolor{black}{To address these challenges, our UA-RIS employs reflection units in Tonpilz design~\cite{lurton2002introduction}, as shown in Fig.~\!\ref{fig:Cir}. Each unit consists of two PZT pads sandwiched between a conical-shaped head mass and a tail mass. The conical-shape of the front mass ensures efficient forward reflection of acoustic waves. The dual-pad push-pull configuration enhances electromechanical coupling and improves energy conversion efficiency between electrical and acoustic signals. Compared to conventional acoustic transducers, these reflection units are smaller and more cost-effective, facilitating the practical, large-scale deployment of UA-RIS.}

\subsection {Hardware Architecture of UA-RIS}
\label{sec:Arc}

\begin{figure}[htb]
\centerline{\includegraphics[width=8.0cm]{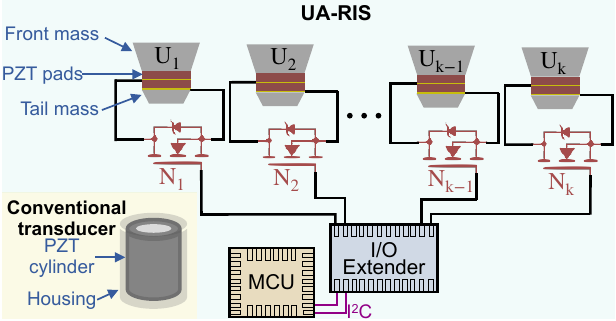}}
  \caption{\textcolor{black}{Hardware architecture of the 1-bit phase coding UA-RIS}. }\label{fig:Cir}
\end{figure}

The architecture of our 1-bit phase coding UA-RIS is shown in Fig.~\!\ref{fig:Cir}.  The load impedance of the reflection unit is \textcolor{black}{dynamically} controlled by the MCU, which switches the associated NMOS transistor ON and OFF to toggle the load impedance between zero (short circuit) and infinity (open circuit). This manipulation allows for adjusting the phase of the reflected wave between 180$^{\circ}$ and 0$^{\circ}$, respectively. \textcolor{black}{To efficiently manage the large array of reflection units on the UA-RIS platform, particularly the 24 units in our prototype, we enhanced the MCU's I/O capabilities using TCA9555 I/O extenders. Each extender is able to control sixteen reflection units.} 

\textcolor{black}{The UA-RIS system operates in three primary modes: phase manipulation, passive reflection, and deep sleep. Phase manipulation is performed at the start of each operational cycle: the MCU regulates the states of NMOS transistors by writing `0' or `1' bitstreams via I$^2$C serial communication. This phase manipulation, which typically lasts a few milliseconds, consumes tens of $\mu$J of energy.
Once phase adjustment is complete, the UA-RIS shifts to a passive reflection state, reflecting incoming signals without generating or amplifying them. Power consumption during this state varies with supply voltages but generally remains on the order of several mW.
In the absence of acoustic communications, the UA-RIS  can enter a deep sleep mode to conserve energy, with power consumption dropping to less than 1 mW~\cite{microchip2015atmega256}. Given these low energy demands, ocean energy harvesters, such as Wirewalkers, which can generate approximately 1 watt of power~\cite{kelly2023prototyping}, are adequate to sustain the UA-RIS's operational requirements.}

\subsection {Beam Patterns of Reflected Waves}
\label{sec:Beam}
By coding the ON/OFF patterns of NMOS transistors, the UA-RIS can manipulate reflected waves to interfere constructively or destructively in designated directions. This capability facilitates the creation of directivity, which significantly enhances or diminishes the signal strength at specific devices. 

\begin{figure}[htb]
\centerline{\includegraphics[width=8.0cm]{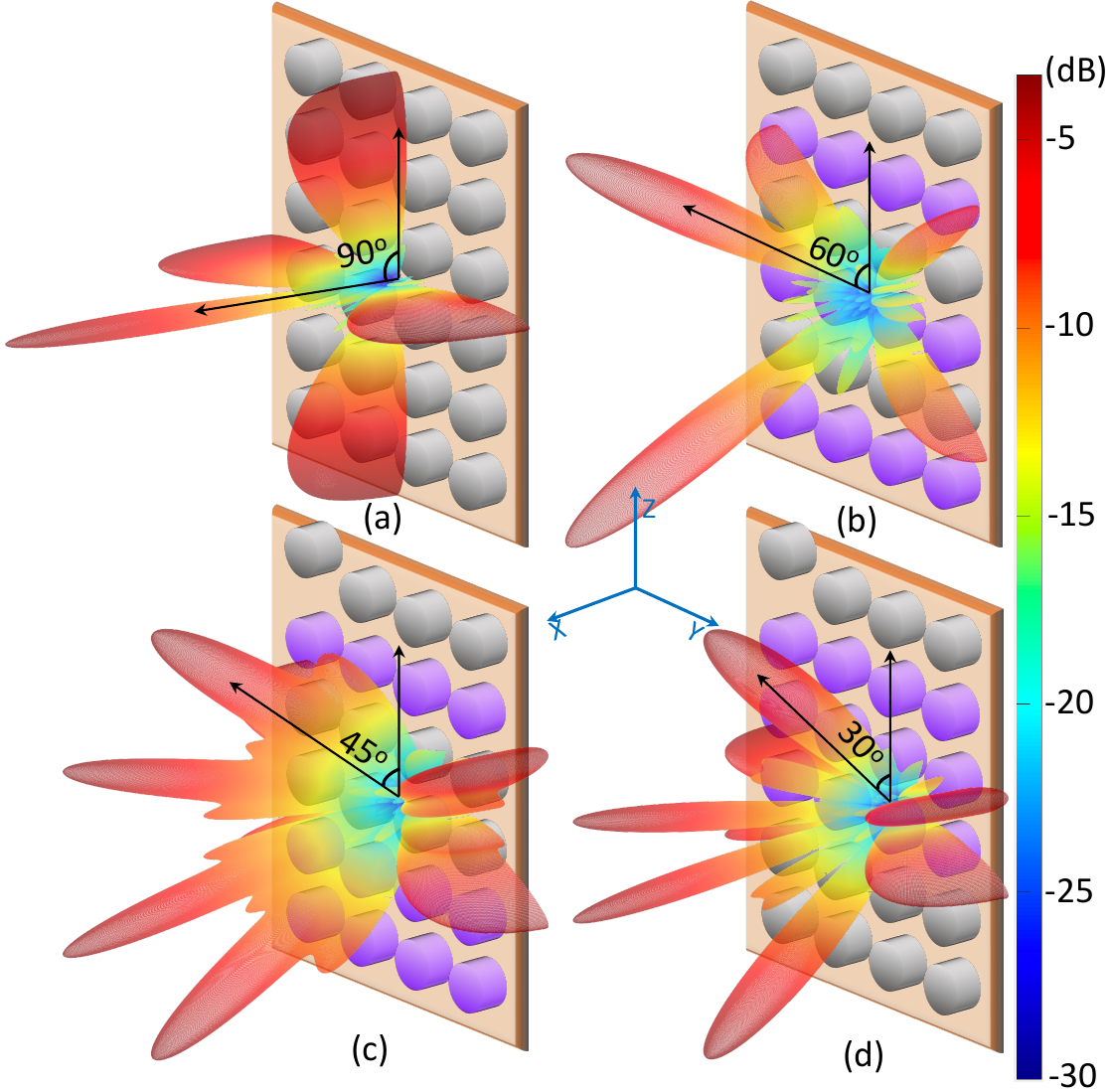}}
  \caption{Beam patterns of reflected waves with different coding schemes.}\label{fig:TankBeam}
\end{figure}

In Fig.~\!\ref{fig:TankBeam}, we present four distinct coding schemes and their respective beam patterns for the reflected waves. The UA-RIS comprises 4$\times$6 reflection units, each separated by a distance of 5\,cm. The UA-RIS  is positioned on the Y-Z plane to reflect a 27\,kHz plane wave traveling along the X-axis. In these diagrams, the ON/OFF patterns of NMOS transistors linked to purple cells are the inverse of those connected to gray cells.

As demonstrated in Fig.~\!\ref{fig:TankBeam}, employing appropriate coding schemes allows for the flexible steering of the main lobes of reflected waves toward any desired direction, across varying zenith and azimuth angles. However, it is important to note that the spacing between neighboring reflection units in the figure exceeds half the wavelength of the incident acoustic wave (i.e., 2.8\,cm at 27\,kHz), leading to spatial aliasing in the reflected wave. Consequently, multiple main lobes may coexist, as observed when the zenith angles are 45$^{\circ}$ or 30$^{\circ}$. This issue can be addressed by reducing the unit spacing.


\section{Experimental Results}
\label{sec:Exp}

This section presents the experimental results obtained from both tank and lake environments, which substantiate the operational viability of the proposed UA-RIS in real applications.

\subsection {Tank Tests}
\label{sec:Tank}

\begin{figure}[htb]
\centerline{\includegraphics[width=8.0cm]{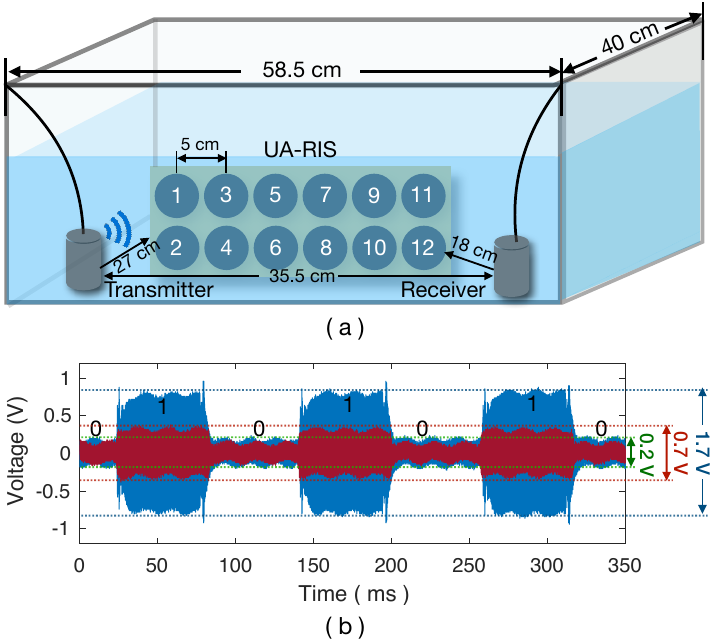}}
  \caption{Tank tests. (a) Experimental settings. (b) Received waveforms, with the blue and  red curves representing the results when 12 and 4 reflection units are operating, respectively.}\label{fig:TankScen}
\end{figure}

The experimental configuration for the tank tests is depicted in Fig.~\!\ref{fig:TankScen}\,(a). Each reflection unit has a resonant frequency of 27\,kHz. The setup includes two omnidirectional underwater acoustic transducers functioning as the transmitter and receiver, respectively. The transmitter is directly connected to a signal generator that emits a 27.13\,kHz sinusoidal signal with a peak-to-peak voltage of 2.0\,V. The receiver is connected to a digital oscilloscope through a preamplifier, enhancing the signal by 46\,dB for effective data collection.

The MCU is programmed to toggle the outputs of I/O pins connected to the gate terminal of the NMOS every 60\,ms. This action switches the loads of reflection units between open circuit and short circuit periodically. Due to the size constraint, only 12 out of 24 reflection units are utilized in the tank tests.

\begin{figure*}[htb]
\vspace{-0.2cm}
\centering
\subfigure []{
\includegraphics[width=3.85cm]{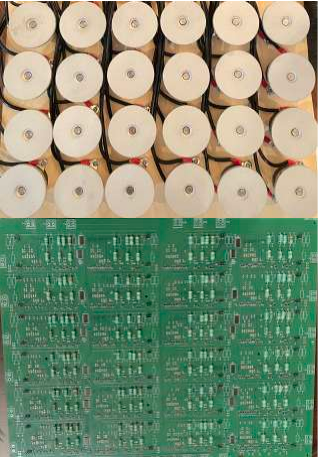}
\label{fig:prototype}
  }
  \hspace{0.1cm}
  \subfigure []{
\includegraphics[width=5.5cm]{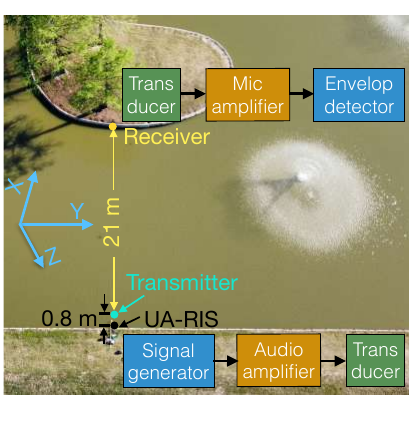}
\label{fig:LakeTest}
}
  \hspace{0.1cm}
  \subfigure []{
\includegraphics[width=5.5cm]{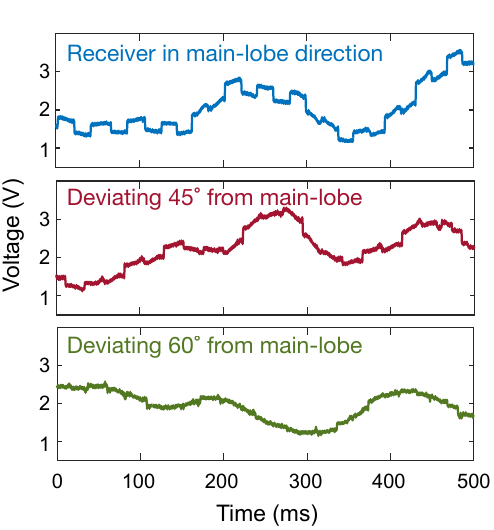}
\label{fig:LakeTestResult}
}
\caption{\textcolor{black}{UA-RIS lake tests. (a) Photo of reflector array (top) and control board (bottom). (b) Experiment setup. (c) Lake test results.}}
\vspace{-0.3cm}
\end{figure*}

In Fig.~\!\ref{fig:TankScen}\,(b), the collected waveforms at the receiver demonstrate the capability of UA-RIS to control the strength of received signals. The blue curve illustrates the enhanced signal strength achieved when 12 reflection units are activated and aligned in phase (code 1). In this configuration, the peak-to-peak voltages ascend to 1.7\,V, as the reflected signals from these units constructively superpose with the direct signal, effectively amplifying the received signal. Conversely, when the phases of all elements are inverted (from 0$^{\circ}$ to 180$^{\circ}$ with code 0), this configuration causes the reflected signals to contract from the direct signal, significantly weakening the received signal to a peak-to-peak voltage of 0.2\,V, which is merely 11.7\% of the voltage observed with code 1.

We also evaluated the performance when only 4 reflection units, specifically units NO.\,9 to NO.\,12 in Fig.~\!\ref{fig:TankScen}\,(a), were activated. The received waveform, shown by the red curve in Fig.~\!\ref{fig:TankScen}\,(b), exhibits a less pronounced change in the strength of the received waves compared to the activation of all 12 units (blue curve). With these four units operating in phase (code 1), the peak-to-peak voltage reaches merely 0.7\,V. This result illustrates the reduced efficacy when fewer units are engaged.

These results confirm that the proposed UA-RIS is adept at manipulating the phase of reflected waves, thus effectively enhancing or diminishing the signal strength at the receiver side. Notably, the system's performance is significantly boosted as the number of operational reflection units is increased, demonstrating the critical role of unit quantity in optimizing the effectiveness of the UA-RIS configuration.

\subsection {Lake Experiments}
\label{sec:Lake}
\textcolor{black}{We conducted lake tests to validate the functionality of our developed UA-RIS system, the photo of which is shown in Fig.~\!\ref{fig:prototype}. Fig.~\!\ref{fig:LakeTest} illustrates the experiment setup.} In the tests, a signal generator produced a 27.76\, kHz continuous wave at -10\,dBm. This signal was subsequently amplified by a 40\,dB audio power amplifier to drive the omnidirectional acoustic transmitter, with no impedance matching. The UA-RIS, consists of 6$\times$4 reflection units, positioned 0.8\,m behind the transmitter on the Y-Z plane. A Microchip ATmega256RFR2 MCU is programmed to toggle the reflection units of the UA-RIS between short circuit and open circuit states every 20\,ms, based on the predefined coding scheme introduced in Fig.~\!\ref{fig:TankBeam}.

At the receiving end, an omnidirectional transducer, placed 21\,m away from the transmitter, is used as receiver. The received wave was first amplified by a microphone amplifier with a 54\,dB gain. Subsequently, the signal was processed through an envelope detector to extract its envelope. Finally, the waveform was recoded by a digital oscilloscope at a sampling rate of 10\,kHz, followed by processing on a laptop.

The experimental results are presented in Fig.~\!\ref{fig:LakeTestResult}, illustrating significant differences in received signal strength based on the coding schemes of the UA-RIS. As shown in the top figure, when the main lobe of the UA-RIS is directed at 90$^{\circ}$ on the Y-Z plane, facing directly towards the receiver, and using coding scheme (a) from Fig.~\ref{fig:TankBeam}, the reflected waves are constructively superimposed \textcolor{black}{with the direct signal with `1' bits and destructively interfered with `0' bits}. This optimal alignment results in the most substantial amplitude change, achieving an average of 280 mV \textcolor{black}{when the MCU toggles the output every 20\,ms}.  In contrast, when the receiver is 45$^{\circ}$ and 60$^{\circ}$ deviating  from the main lobe and UA-RIS employing coding schemes (c) and (d) from Fig.~\!\ref{fig:TankBeam} respectively, the voltage changes observed in the received waves are only 100\,mV and 33\,mV, respectively. 

\textcolor{black}{It is important to note that the measured amplitude differences do not precisely represent the strength of the reflected signal, as the envelope detector at the receiver compresses the amplitude of the signal nonlinearly. The amplitude differences presented in Fig.~\!\ref{fig:LakeTestResult} qualitatively demonstrate the influence of UA-RIS orientation on the effectiveness of signal enhancement, which can be adjusted using different phase coding schemes. This provides a basic yet insightful view into how UA-RIS technology can be utilized to enhance underwater communication systems.}

\subsection{Discussions}
\textcolor{black}{While the proposed UA-RIS architecture demonstrates feasibility for manipulating the strength of the acoustic waves at target directions, our lake experiments face challenges constrained by the current resources and more comprehensive experiments are necessary to fully understand its potential in enabling high-rate, long-range acoustic communications.} 

\vspace{0.1cm}
\begin{adjustwidth}{-0.77cm}{0cm}
\begin{description}
\setlength{\labelsep}{-0.95em}
\itemsep 0.07cm
  \item[a)] \textcolor{black}{\textbf{Quantitative Performance Analysis of UA-RIS:}  Accurate measurement of the directional gain achieved with the UA-RIS and the assessment of the performance gap between experimental results and theoretical predictions are crucial. Such measurements require an \emph{anechoic tank} that eliminates reflections, scattering, and environmental factors like wind and waves. Additionally, employing well-calibrated standard hydrophones precisely installed at the same distance but in different directions around the UA-RIS is essential. Conducting these controlled experiments will provide valuable insights into the system's performance and validate its effectiveness.}

\textcolor{black}{However, due to resource constraints, the ``Receiver in main-lobe direction'' test shown in Fig.~\!\ref{fig:LakeTestResult} was conducted by visually aligning the receiver and monitoring the envelope on an oscilloscope. This method, while practical, introduces visual errors that can lead to deviations from the optimal main-lobe direction.  Additionally, in the experiments, the transmitter, receiver, and UA-RIS were not rigidly fixed but hung on the lake bank. This arrangement made the system susceptible to slight positional changes due to wave action, which in turn impacted performance.  }

  \item[b)] \textcolor{black}{\textbf{Addressing Hardware Assembly Errors:} Inaccurate spacing and installation of the reflection units had a significant impact on the real-world performance of the UA-RIS. During the assembly process, variations of up to 3\% were observed in the spacing. Moreover, inconsistencies in the assembly process caused deviations in the orientation of the reflection units. These deviations became more pronounced over time due to the weight of the front mass. Unlike the lightweight, printable antennas used in RF-RIS, the reflection units in UA-RIS are considerably larger and heavier, making the installation far more challenging.} 

\textcolor{black}{During lake experiments, we observed that the reflection units contributed unevenly, leading to a directional gain significantly lower than expectation. While hardware imperfections were suspected, other installation-related factors also likely contributed and require further investigation. Future research will aim to achieve ideal hardware conditions for higher system efficiency. Alternatively, characterizing hardware imperfections and incorporating them into theoretical models can help mitigate the performance gap between experimental results and theoretical analysis. This approach will enhance the understanding of how physical deviations affect system performance and guide improvements in design and assembly processes.}

  \item[c)] \textcolor{black}{\textbf{Large-Scale Deployment and Exploration of High-Rate Communication:} More comprehensive testing in practical experimental settings is required to evaluate the UA-RIS's efficiency in extending the communication range of high-rate acoustic communications. Practical deployments in diverse operating conditions will allow for the comprehensive assessment of the system's performance in real-world applications. 
  In addition, extending the operational frequency range of the UA-RIS to over 100\,kHz is an important direction for future research. Higher frequencies can support increased data rates, which are essential for applications requiring high-rate underwater communication such as in AUV networks. This extension will involve redesigning reflection units and optimizing system components to handle higher frequencies effectively. }
  
\end{description}
\end{adjustwidth}
\vspace{0.1cm}


\section{Conclusion and Future Work}
\label{sec:Con}
\textcolor{black}{In this article, we proposed a cost-effective design for UA-RIS that addresses the unique challenges of underwater environments. Instead of relying on varactor-based phase manipulation -- which is ineffective in acoustic frequency due to insufficient tuning ranges -- our UA-RIS design employs a load network that achieves flexible phase shifts and high energy efficiency. A 1-bit phase coding based UA-RIS prototype consisting of 24 reflection units was developed and validated through  tank and lake experiments. The experimental findings indicate that the developed UA-RIS can effectively manipulate the strength of the acoustic waves at target direction. }

\textcolor{black}{UA-RIS is still in its nascent stages. The practical adoption of UA-RIS assisted networks still requires significant development. For instance, to support more advanced beamforming coding schemes, it is essential to move beyond the current 1-bit phase coding. Our short-term research goal is to develop UA-RIS platform capable of producing reflected waves with arbitrary phase shifts and amplitude manipulation.
Additionally, improving the resilience of the control board to handle a wide range of incident signal intensities and high variation of transducer impedance in dynamic underwater environments are being investigated. 
To ensure communication efficiency in UA-RIS integrated systems, modifications to existing medium access control protocols are necessary. Adapting these protocols will facilitate seamless integration of UA-RIS into existing  underwater network infrastructures, which will be investigated in our long-term research.}

\bibliographystyle{IEEEtran}
\bibliography{UnderBack}

\end{document}